\documentclass{ipres_proc_article-sp}

\usepackage{subfigure}
\usepackage{graphicx}

\usepackage{url}
\usepackage{amsmath}
\usepackage{multicol}
\usepackage{fancyvrb}
\usepackage{float}
\usepackage{xcolor}
\usepackage{array}
\usepackage{multirow}
\usepackage{tabularx}
\usepackage{colortbl}
\usepackage{hhline}
\usepackage{spverbatim}
\usepackage{listings}
\usepackage[T1]{fontenc}
\usepackage[utf8]{inputenc}
\usepackage{lmodern}

\def\firstitem#1{\addtolength{\itemsep}{-0.5\baselineskip} 
\item{\vskip -.10in #1}}

\usepackage[sort&compress, square, comma, numbers]{natbib}

\begin{document}

\title{Archiving Deferred Representations Using a Two-Tiered Crawling Approach}

\numberofauthors{1}

\author{
\alignauthor
Justin F. Brunelle, Michele C. Weigle, and Michael L. Nelson\\
       \affaddr{Old Dominion University}\\
       \affaddr{Department of Computer Science}\\
       \affaddr{Norfolk, Virginia, 23508}\\
       \email{\{jbrunelle, mweigle, mln\}@cs.odu.edu}
}

\maketitle
\begin{abstract}
Web resources are increasingly interactive, resulting in resources that are increasingly difficult to archive. The archival difficulty is based on the use of client-side technologies (e.g., JavaScript) to change the client-side state of a representation after it has initially loaded. We refer to these representations as \emph{deferred representations}. We can better archive deferred representations using tools like headless browsing clients. 
We use 10,000 seed Universal Resource Identifiers (URIs) to explore the impact of including PhantomJS -- a headless browsing tool -- into the crawling process by comparing the performance of wget (the baseline), PhantomJS, and Heritrix. Heritrix crawled 2.065 URIs per second, 12.15 times faster than PhantomJS and 2.4 times faster than wget. However, PhantomJS discovered 531,484 URIs, 1.75 times more than Heritrix and 4.11 times more than wget. To take advantage of the performance benefits of Heritrix and the URI discovery of PhantomJS, we recommend a tiered crawling strategy in which a classifier predicts whether a representation will be deferred or not, and only resources with deferred representations are crawled with PhantomJS while resources without deferred representations are crawled with Heritrix. We show that this approach is 5.2 times faster than using only PhantomJS and creates a frontier (set of URIs to be crawled) 1.8 times larger than using only Heritrix.
\end{abstract}




\section{Introduction}
\label{introduction}
The Web -- by design and demand -- continues to change. Today's Web users expect Web resources to provide application-like interactive features, client-side state changes, and personalized representations. These features enhance the browsing experience, but make archiving the resulting representations difficult -- if not impossible. We refer to the ease of archiving a Web resource as \emph{archivability} \cite{ijdl}. 


Web resources are ephemeral by nature, making archives like the Internet Archive \cite{iawebarchive, waybackarchives2} valuable to Web users seeking to revisit prior versions of the Web. Users (and robots) utilize archives in a variety of ways \cite{usingIA, marshalls_social_media_study, yasminLinks}. Live Web resources are more heavily leveraging JavaScript (i.e., Ajax) to load embedded resources, which leads to the live Web ``leaking'' into the archive \cite{zombies} or missing embedded resources in the archives, both of which ultimately results in reduced archival quality \cite{brunelleDamage}.

We define we define \emph{deferred representations} as 
representations of resources that use of JavaScript and other client-side technologies
to load embedded resources or fully construct a representation and, therefore, have
low archivability. \emph{Deferred} refers to the final representation that is not fully realized
and constructed until after the client loads the page and executes the client-side
code. The client will render the representation on the user-agent and user interactions and events that occur within the
representation on the client. The final representation is deferred until after the user-agent, JavaScript, and user events complete their execution 
on the resource. From this point forward, we will refer to representations dependent
upon these factors as deferred representations.

Conventional Web crawlers (e.g., Heritrix, wget) are not equipped with the necessary tools to execute JavaScript during the archival process \cite{googleJS} and subsequently never dereference the URIs of the resources embedded via JavaScript and are required to complete the deferred representation. PhantomJS allows JavaScript to execute on the client, rendering the representation as would a Web browser. In the archives, the missing embedded resources return a non-200 HTTP status (e.g., 404, 503) when their Universal Resource Identifiers (URIs) are dereferenced, leaving pages \emph{incomplete}. Deferred representations can also lead to \emph{zombies} which occur when archived versions of pages inappropriately load embedded resources from the live Web \cite{zombies}, leaving pages incorrect, or more accurately, \emph{prima facie violative} \cite{ainsworthTR}. 

We investigate the impact of crawling deferred representations as the first step in an improved archival framework that can replay deferred representations both completely and correctly. We measure the expected increase in frontier (list of URIs to be crawled) size and wall-clock time required to archive resources, and investigate a way to recognize deferred representations to optimize crawler performance using a two-tiered approach that combines PhantomJS and Heritrix. Our efforts measure the crawling tradeoff between traditional archival tools and tools that can better archive JavaScript with headless browsing -- a tradeoff that was anecdotally understood but not yet measured.

Throughout this paper we use Memento Framework terminology. Memento \cite{nelson:memento:tr} is a framework that standardizes Web archive access and terminology. Original (or live web) resources are identified by URI-R, and archived versions of URI-Rs are called \emph{mementos} and are identified by URI-M. 

\section{Related Work}
\label{priorwork}
Archivability helps us understand what makes representations easier or harder to archive. Banos et al. created an algorithm to evaluate archival success based on adherence to standards for the purpose of assigning an archivability score \cite{ipresArchivability}. In our previous work, we studied the factors influencing archivability, including accessibility standards and their impact on memento completeness, demonstrating that deviation from accessibility standards leads to reduced archivability \cite{kellyTPDL2013}. We also demonstrated the correlation between the adoption of JavaScript and Ajax and the number of missing embedded resources in the archives \cite{ijdl}. 

Spaniol measured the quality of Web archives based on matching crawler strategies with resource change rates \cite{spaniol9catch, spaniol2009data, Denev:2009:SFQ:1687627.1687694}. Ben Saad and Gan\c{c}arski performed a similar study regarding the importance of changes on a page \cite{saad2011}. Gray and Martin created a framework for high quality mementos and assessed their quality by measuring the missing embedded resources \cite{mementoQuality}. In previous work, we measured the relative damage caused to mementos that were missing embedded resources to quantify the damage caused by missing resources loaded by JavaScript \cite{brunelleDamage}. These works study quality, helping us understand what is missing from mementos.

David Rosenthal spoke about the difficulty of archiving representations enabled by JavaScript \cite{iipc2013, futureWeb}. Google has made efforts toward indexing deferred representations -- a step in the direction of solving the archival challenges posed by deferred representations \cite{googleJS}. Google's indexing focuses on rendering an accurate representation for indexing and discovering new URIs, but does not completely solve the challenges to archiving caused by JavaScript. Archiving web resources and indexing representation content are different activities that have differing goals and processes.

Several efforts have studied client-side state. Mesbah et al. performed several experiments regarding crawling and indexing representations of Web pages that rely on JavaScript \cite{mesbahCrawling, mesbahInferState}. These works have focused mainly on search engine indexing and automatic testing \cite{mesbahTesting, mesbah2} rather than archiving, but serve to illustrate the pervasive problem of deferred representations. Dincturk et al. constructed a model for crawling Rich Internet Applications (RIAs) by discovering all possible client-side states and identifying the simplest possible state machine to represent the states \cite{dincturkAjax}. 

These prior works have focused on archival difficulties of crawling and indexing deferred representations, but have not explored the impact of archiving deferred representations on archival processes and crawlers. We measure the trade-off between speed and completeness of crawling techniques.
\\
\\
\section{Background}
\label{backgroung}
Web crawlers operate by starting with a finite set of seed URI-Rs in a frontier -- or list of crawl targets -- and add to the frontier by extracting URIs from the representations returned. 
Representations of Web resources are increasingly reliant on JavaScript and other client-side technologies to load embedded resources and control the activity on the client. Web browsers use a JavaScript engine to execute the client side code; Web crawlers traditionally do not have such an engine or the ability to execute client-side code because of the resulting loss of crawling speed. The client-side code can be used to request additional data or resources from servers (e.g., via Ajax) after the initial page load. Crawlers are unable to discover the resources requested via Ajax and, therefore, are not adding these URIs to their frontiers. The crawlers are missing embedded resources, which ultimately causes the mementos to be incomplete. 

To mitigate the impact of JavaScript and Ajax on archivability, traditional crawlers that do not execute JavaScript (e.g., Heritrix) have constructed approaches for extracting links from embedded JavaScript to be added to crawl frontiers. Even though it does not execute JavaScript, Heritrix v. 3.1.4 does peek into the embedded JavaScript code to extract links where possible \cite{htrixJS}. These processes rely on string matching and regular expressions to recognize URIs mentioned in the JavaScript. This is a sub-optimal approach because JavaScript may construct URIs from multiple strings during execution, leading to an incomplete URI extracted by the crawler.  

Because archival crawlers do not execute JavaScript, what is archived by automatic crawlers is increasingly different than what users experience. A solution to this challenge of archiving deferred representations is to provide crawlers with a JavaScript engine and allow headless browsing (i.e., allow a crawler to operate like a browser) using a technology such as PhantomJS. However, this change in crawling method impacts crawler performance, frontier size, and crawl time.

\section{Motivating Examples}
\label{example}

\begin{figure*}
  \begin{center}
    \subfigure[The live resource at URI-R \protect\url{http://www.truthinshredding.com/} loads A, B, and C via JavaScript.]{\label{liveex}\includegraphics[width=0.3\textwidth]{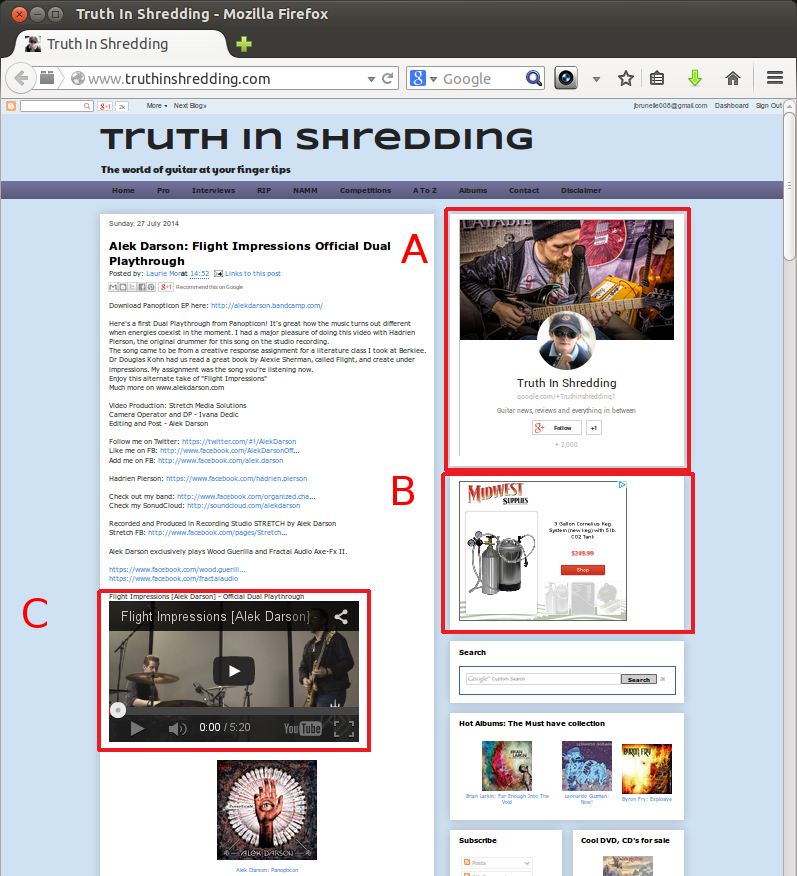}}\qquad
    \subfigure[Using PhantomJS, the advertisement (B) and video (C) are found but the account frame (A) is missed.]{\label{pjsex}\includegraphics[width=0.3\textwidth]{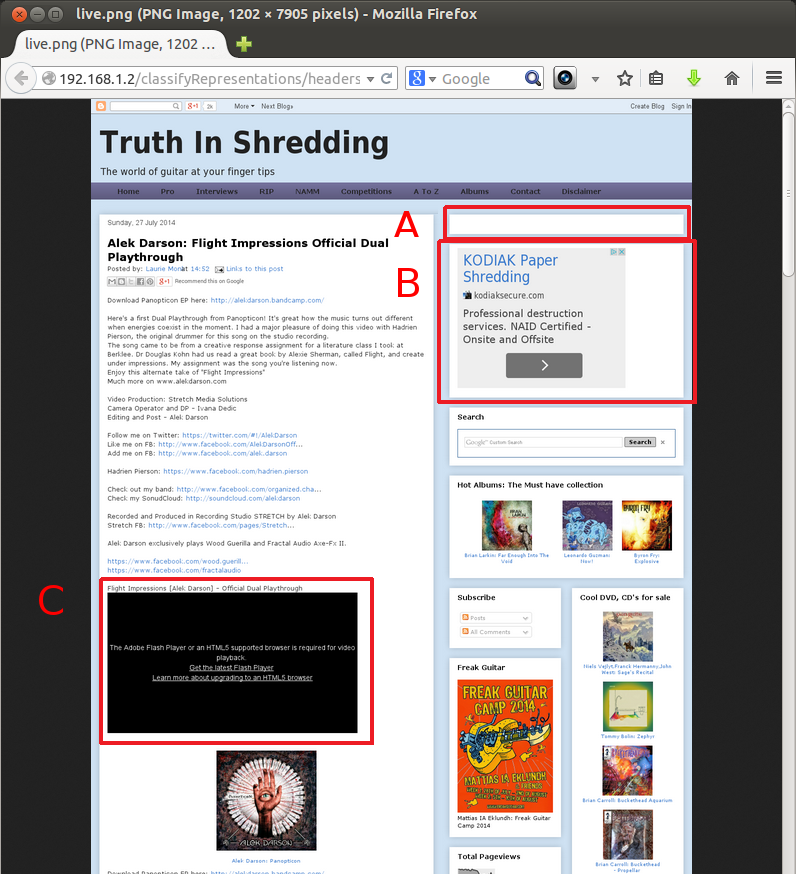}}  \qquad
    \subfigure[Using Heritrix, the embedded resources A, B, and C are missed.]{\label{wayex}\includegraphics[width=0.3\textwidth]{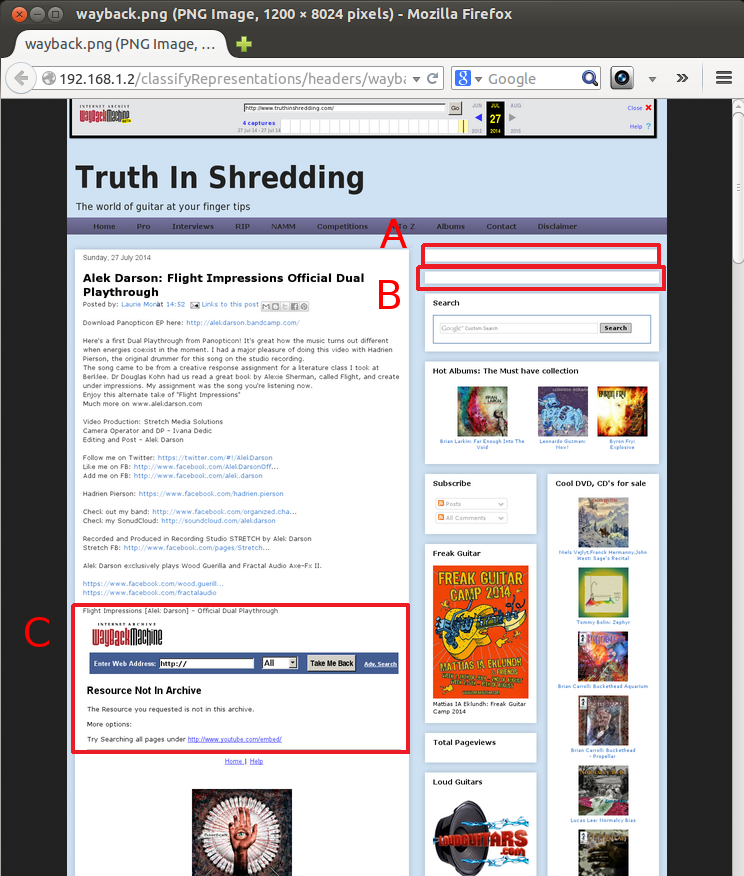}}  \qquad
  \end{center}
\vskip -3mm
  \caption{Neither archival tool captures all embedded resources, but PhantomJS discovers the URI-Rs of two out of three embedded resources dependent upon JavaScript (B, C) while Heritrix misses all of them.}
  \label{motivate}
\end{figure*}

To illustrate the challenge of archiving resources with deferred representations, we consider the resource at URI-R \url{http://www.truthinshredding.com/} and its mementos in Figure \ref{motivate}. We took a PNG snapshot of the live-Web resource as rendered in Mozilla Firefox (Figure \ref{liveex}), the resource as loaded by PhantomJS (Figure \ref{pjsex}), and the memento created by Heritrix and viewed in a local installation of the Wayback Machine (Figure \ref{wayex}). The title of the page ``Truth in Shredding'' appears in a different font in Figure \ref{liveex} than in Figures \ref{pjsex} and \ref{wayex} not due to a missing style sheet but rather an incompatibility of the font for the headless browser. 

The live-Web resource loads embedded resources (annotated as A, B, and C) via JavaScript. Embedded Resource A is an HTML page loaded into an iframe. The original resources are described in Table \ref{urirs}.

\begin{table*}
\centering
\begin{tabular}{l|p{15cm}}
URI ID & URI-R\\
\hline          
\hline          
$URI$-$R_{A}$ & \url{https://apis.google.com/u/0/_/widget/render/page?usegapi=1&rel=publisher&href=%2F%2Fplus.google.com%2F110743665890542265089&width=430&hl=en-GB&origin=http%3A%2F%2Fwww.truthinshredding.com&gsrc=3p&ic=1&jsh=m%3B%2F_%2Fscs%2Fapps-static...}\\
\hline          
$URI$-$R_{A_1}$ & \url{https://apis.google.com/_/scs/apps-static/_/ss/k=oz.widget.-ynlzpp4csh.L.W.O/m=bdg/am=AAAAAJAwAA4/d=1/rs=AItRSTNrapszOr4y_tKMA1hZh6JM-g1haQ}\\
\hline            
$URI$-$R_{B_1}$  & \url{http://pagead2.googlesyndication.com/pagead/show_ads.js}\\
\hline                   
$URI$-$R_{C}$ & \url{http://www.youtube.com/embed/QyLl4Fd4cGA?rel=0}\\
\hline
\end{tabular}
\caption{The URI-Rs in Figure \ref{motivate}.}
\label{urirs}
\end{table*}

Embedded Resource A, after it loads into the iframe, uses JavaScript to pull the profile image into the page from $URI$-$R_{A_1}$. Embedded Resource B is an advertisement that uses the JavaScript at $URI$-$R_{B_1}$ 
to pull in ads to the page. Embedded Resource C is a YouTube video that is embedded in the page using the following HTML for an iframe:
 
\begin{verbatim}
<iframe allowfullscreen="" frameborder="0" height=
"281" src="//www.youtube.com/embed/QyLl4Fd4cGA?rel
=0" width="500"></iframe>.
\end{verbatim}

PhantomJS does not load Embedded Resource A, potentially because the host resource completes loading before the page embedded in the iframe can finish loading. PhantomJS stops recording embedded URIs and monitoring the representation after a page has completed loading, and Embedded Resource A executes its JavaScript to load the profile picture after the main representation has completed the page load\footnote{PhantomJS scripts can be written to avoid this race-condition using longer timeouts or client-side event detection, but this is outside the scope of this paper.}.
 PhantomJS does discover the advertisement (Embedded Resource B) and the YouTube video (Embedded Resource C). Even though the headless browser used by PhantomJS does not have the plugin necessary to display the video, the URI-R is still discovered by PhantomJS.

Heritrix fails to identify the URI-Rs for the Embedded Resources A, B, and C. When the memento created by Heritrix is loaded by the Wayback Machine, Embedded Resources A, B, and C are missing. This is attributed to Heritrix, which does not discover the URI-Rs for these resources during the crawl. When viewing the memento through the Wayback Machine, the JavaScript responsible for loading the embedded resources is executed resulting in either a zombie resource (\emph{prima facie violative}) or HTTP 404 response (incomplete) for the embedded URI.

Heritrix's inability to discover the embedded URI-Rs could be mitigated by utilizing PhantomJS during the crawl. However, this raises many questions, most notably: How much slower will the crawl time be? How many additional embedded resources could it recover and potentially need to store? Can we optimize the crawl approach based on the detection of deferred representations? Our investigation into these questions will assess the feasibility of combining Heritrix with PhantomJS to balance the speed of Heritrix with the completeness of PhantomJS.

\section{Comparing Crawls}
\label{performance}
We designed an experiment to measure the performance differences between a command-line archival tool (wget \cite{wget}), a traditional crawler (the Internet Archive's Heritrix Crawler \cite{heritrix, Sigurosson:Incremental-Heritrix}), and a headless browser client (PhantomJS). Neither Heritrix nor wget execute the client-side JavaScript, while PhantomJS \emph{does} execute client-side JavaScript.

We constructed a 10,000 URI-R dataset by randomly generating a Bitly URI and extracting its redirection target (identical to the process used to create the Bitly data subset in \cite{hmotwia}). We split the 10,000 URI dataset into 20 sets of 500 seed URI-Rs and used wget, Heritrix, and PhantomJS to crawl each set of seed URI-Rs. We repeated each crawl ten times to establish an average performance, resulting in ten different crawls of the 10,000 URI dataset (executing the crawl one of the 500-URI sets at a time) with wget, Heritrix, and PhantomJS. We measured the increase in frontier size ($|F|$) and the URIs per second ($t_{URI}$) to crawl the resource. 

While Heritrix provides a user interface that identifies the crawl frontier size, PhantomJS and wget do not. We calculate the frontier size of PhantomJS by counting the number of embedded resources that PhantomJS requests when rendering the representation. We calculate the frontier size of wget by executing a command\footnote{We executed \texttt{wget -T 40 -o outfile -p -O headerFile [URI-R]} which downloads the target URI-R and all embedded resources and dumps the HTTP traffic to \texttt{headerFile}.} that records the HTTP GET requests issued by wget during the process of mirroring a web resource and its embedded resources. We consider the frontier size to be the total number of resources and embedded resources that wget attempts to download.

We began a crawl of the same 500 URI-Rs using wget, Heritrix, and PhantomJS simultaneously to mitigate the impact of live Web resources changing state during the crawls. For example, if the representation changes (such as includes new embedded resources) in between the times wget, PhantomJS, and Heritrix perform their crawls, the number or representations of embedded resources may change and therefore the representation influenced the crawl performance, not the crawler itself. 

We crawled live-Web resources because mementos inherit the limitations of the crawler used to create them. Depending on crawl policies, a memento may be incomplete and different than the live resource. The robots.txt protocol \cite{robots, robotsProtocol}, breadth- versus depth-first crawling, or the inability to crawl certain representations (like deferred representations as we discuss in this paper) can all influence the mementos created during a crawl.


\subsection{Crawl Time by URI}
\label{clock}
\begin{figure}
  \begin{center}
    	\includegraphics[width=0.45\textwidth,keepaspectratio]{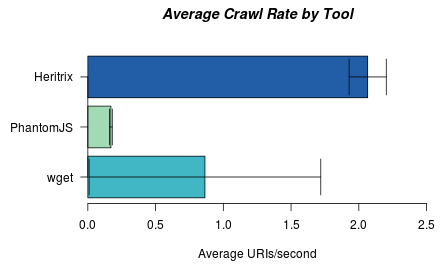}
  \end{center}
  \caption{Heritrix crawls 12.13 times faster than PhantomJS. The error lines indicate the standard deviation across all ten runs.}
  \label{crawlrate}
\end{figure}

To better understand how crawl times of wget, PhantomJS, and Heritrix differ, we determined the time needed to execute a crawl. Heritrix has a browser-based user interface that provides the URIs/second ($t_{URI}$) metric. We collected this metric from the Web interface for each crawl. We used Unix system times to calculate the crawl time for each PhantomJS and wget crawl by determining the start and stop times for dereferencing each resource and its embedded resources. We compare the wget, PhantomJS, and Heritrix crawl times per URI in Figure \ref{crawlrate} and Table \ref{perfTable}. Heritrix outperforms PhantomJS, crawling 2.065 URIs/s while PhantomJS crawls 0.170 URIs/s and wget crawls 0.864 URIs/s. Heritrix crawls, on average, 12.13 times faster than PhantomJS and 2.39 times faster than wget.

The performance difference comes from two aspects of the crawl. First, Heritrix executes crawls in parallel with multiple threads being managed by the Heritrix software -- this is not possible with PhantomJS on a single core machine since PhantomJS requires access to a headless browser and its associated JavaScript engine, and parallelization will result in process and threading conflicts. Second, Heritrix does not execute the client-side JavaScript and only adds URIs that are extracted from the Document Object Model (DOM), embedded style sheets, and other resources to its frontier. 

\subsection{URI Discovery and Frontier Size}
\label{frontier}
We performed a string-matching de-duplication (that is, removing duplicate URIs) to determine the true frontier size ($|F|$). 

\begin{table}[h]
\begin{tabular}{l|r|r|r|r}
\multicolumn{1}{c}{\multirow{2}{*}{Crawler}} & \multicolumn{2}{c}{\begin{tabular}[c]{@{}c@{}}Crawl time\end{tabular}} & \multicolumn{2}{c}{\begin{tabular}[c]{@{}c@{}}Frontier Size\end{tabular}} \\
\multicolumn{1}{c}{}                         & $\overline{t_{URI}}$                                   & $s_{t_{URI}}$                                  & $\overline{|F|}$                                   & $s_{|F|}$                                 \\
\hline
\hline
wget                                         & 0.864                                   & 0.855                                     & 129,443                                 & 3,213.65                                 \\
\hline
Heritrix                                     & 2.065                                   & 0.137                                     & 302,961                                 & 1,219.82                                 \\
\hline
PhantomJS                                    & 0.170                                   & 0.001                                     & 531,484                                 & 2,036.92     \\        
\hline                   
\end{tabular}
\caption{Mean and standard deviation of crawl time (URIs/s) and frontier size for wget, Heritrix, and PhantomJS crawls of 10,000 seed URIs.}
\label{perfTable}
\end{table}

\begin{figure}
  \begin{center}
    \includegraphics[width=0.45\textwidth]{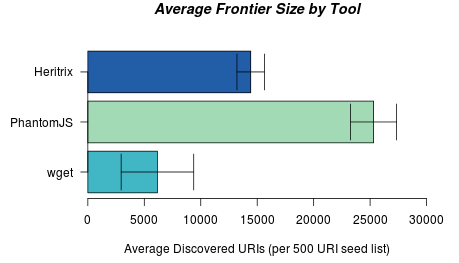}
  \end{center}
  \caption{PhantomJS discovers 1.75 times more embedded resources than Heritrix and 4.11 times more resources than wget. The averages and error lines indicate the standard deviation across all ten runs.}
  \label{frontiersize}
\end{figure}

As shown in Figure \ref{frontiersize} and in Table \ref{perfTable}, we found that PhantomJS discovered and added 
1.75 times more URI-Rs to its frontier than Heritrix, and 4.11 times more URI-Rs than wget. Per URI-R, PhantomJS loads 19.7 more embedded resources than Heritrix and 32.4 more embedded resources than wget. The superior PhantomJS frontier size is attributed to its ability to execute JavaScript and discover URIs constructed and requested by the client-side scripts.

\begin{figure}
  \begin{center}
     \subfigure[A portion of Heritrix, PhantomJS, and wget frontiers overlap. PhantomJS and Heritrix identify URIs that the others do not.]{\label{unionInt}\includegraphics[width=0.48\textwidth]{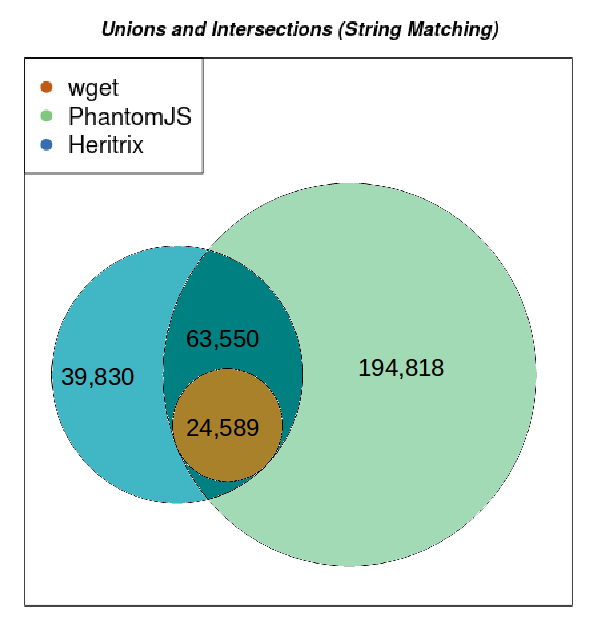}}\\
    \subfigure[The frontier of URI-Rs unique to PhantomJS shrinks when only considering the host and path aspects (Base Policy for matching) of the URI-R.]{\label{fuzzy}\includegraphics[width=0.48\textwidth]{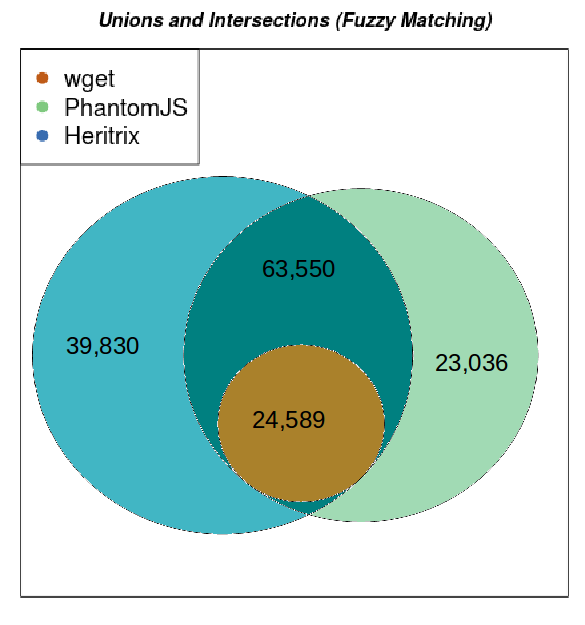}}
  \end{center}
  \caption{Heritrix, PhantomJS, and wget frontiers as an Euler Diagram. The overlap changes depending on how duplicate URIs are identified.}
  \label{vennDiagrams}
\end{figure}

%

\begin{figure}
 \begin{center}
    \includegraphics[width=0.48\textwidth,keepaspectratio]{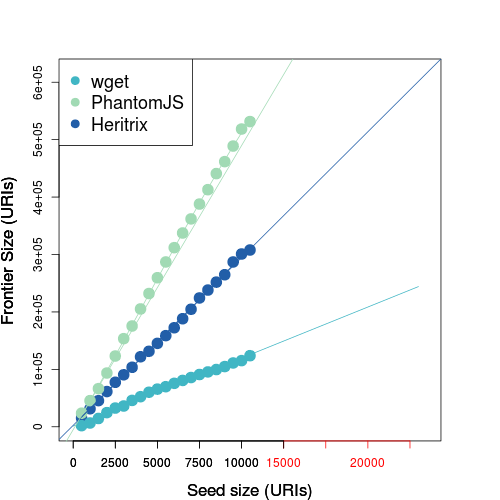}
  \end{center}
  \caption{Frontier size grows linearly with seed size.}
  \label{frontierGrowth}
\end{figure}

However, raw frontier size is not the only performance metric for assessing the quality of the frontier. PhantomJS and Heritrix discover some of the same URIs, while PhantomJS discovers URIs that Heritrix does not and Heritrix discovers URIs that PhantomJS does not. We measured the union and intersection of the Heritrix and PhantomJS frontiers. As shown in Figure \ref{unionInt}, per 10,000 URI-R crawl Heritrix finds 39,830 URI-Rs missed by PhantomJS on average, while PhantomJS finds 194,818 URI-Rs missed by Heritrix per crawl on average. PhantomJS and Heritrix find 63,550 URI-Rs in common between the two crawlers. The wget crawl resulted in a frontier of 24,589 URI-Rs, which was a proper subset of both the Heritrix and PhantomJS frontiers.

\begin{figure}[t]
 \begin{center}
    \includegraphics[width=0.48\textwidth,keepaspectratio]{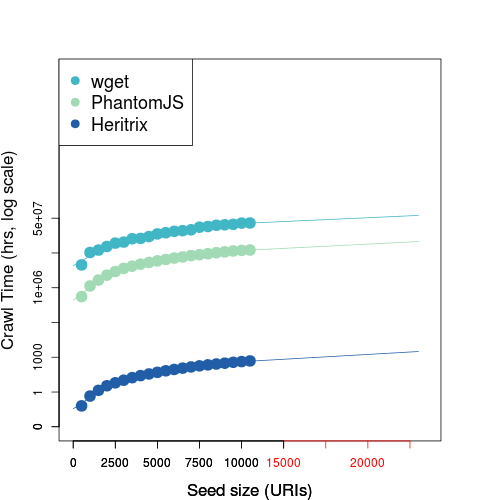}
  \end{center}
  \caption{Crawl speed is dependent upon frontier size.}
  \label{speedGrowth}
\end{figure}

This analysis shows that PhantomJS finds 19.70 more embedded resources per URI than Heritrix (Figure \ref{frontierGrowth}). Heritrix runs 12.13 times faster than PhantomJS (Figure \ref{speedGrowth}). Note that the red axis in Figures \ref{frontierGrowth} and \ref{speedGrowth} are unmeasured and only projections of the measured trends, with the projections predicting the performance as the seed list size grows.

\vskip -3mm

\subsection{Frontier Properties}
\label{frontierProp}
During the PhantomJS crawls, we observed that PhantomJS discovers session-specific URI-Rs that Heritrix misses and Heritrix discovers Top Level Domains (TLDs) that PhantomJS misses, presumably from Heritrix's inspection of JavaScript. For example:  
\begin{verbatim}
http://dg.specificclick.net/?y=3&t=h&u=http%3A%2F%2
Fmisscellania.blogspot.com%2Fstorage%2F
Twitter-2.png...
\end{verbatim}
\noindent from PhantomJS versus
\begin{verbatim}
http://dg.specificclick.net/
\end{verbatim}

\noindent from Heritrix. The uniquely Heritrix URI-Rs are potentially the base of a URI to be further built by JavaScript.  Because PhantomJS only discovers URIs for which the client issues HTTP requests, this URI-R is not discovered by PhantomJS. To determine the nature of the differences between the Heritrix and PhantomJS frontiers, we analyzed the union and intersection between the URI-Rs in the frontiers using different matching policies (Figure \ref{fuzzy}). 

During a crawl of 500 URI-Rs by PhantomJS, 19,022 URI-Rs were added to the frontier for a total of 19,522 URI-Rs in the frontier. We also captured the content body (the returned entity received when dereferencing a URI-R from the frontier) and recorded its MD5 hash value. We used the hash value to identify duplicate representations during the crawl. To determine duplication between URIs, we used five matching policies to determine the duplication within the frontier (Table \ref{dupesTable}). In other words, we identify cases in which the URIs are different but the content is the same, similar to the methods used by Sigurðsson \cite{uriDupes, uriDupes2}.

The \emph{No Trim} policy uses strict string matching of the URI-Rs to detect duplicates. The \emph{Base Trim} policy trims all parameters from the URI. 
The \emph{Origin Trim} policy eliminates all parameters and associated values that reference a referring source, such as \texttt{origin}, \texttt{callback}, \texttt{domain}, or \texttt{referrer}. These parameters are often associated with a value including the top level domain of the referring page. Frequent implementers include Google Analytics or ad services.

The \emph{Session Trim} policy eliminates all parameters and their associated values that reference a session. 
For example, the parameters such as \texttt{session}, \texttt{sessionid}, \texttt{token\_id}, etc. are all removed from the URI-R before matching. These parameters are often used by ad services or streaming media services to identify browsing sessions for tracking and revenue generation purposes.

The \emph{HTTP Trim} policy removes all parameters with values that mention a URI. Ad services, JavaScript files, and other statistics tracking services frequently utilize these parameters. 
Examples of each trim policy are shown in Table \ref{trimTable}.

\begin{table*}
\centering
\begin{tabular}{l|p{6cm}|p{6cm}}
Trim Policy & Original URI-R & Trimmed URI-R\\
\hline          
\hline          
No Trim & \url{http://example.com/folder/index.html?param=value} & \url{http://example.com/folder/index.html?param=value} \\
\hline          
Origin Trim & \url{http://example.com/folder/index.html?callback=cs.odu.edu} & \url{http://example.com/folder/index.html} \\
\hline          
Base Trim & \url{http://example.com/folder/index.html?param=value} & \url{http://example.com/folder/index.html} \\
\hline          
Session Trim & \url{http://example.com/folder/index.html?param=value&sessionid=12345} & \url{http://example.com/folder/index.html?param=value} \\
\hline        
HTTP Trim & \url{http://example.com/folder/index.html?param=value&httpParam=http://www.test.com/} & \url{http://example.com/folder/index.html?param=value} \\
\hline
\end{tabular}
\caption{Examples of the URI trim policies.}
\label{trimTable}
\end{table*}

We used the five trimming policies to detect duplicates in the frontiers constructed by PhantomJS in one of the crawls of 500 URI-Rs. At the end of the crawl, PhantomJS had a frontier of 19,522 URI-Rs. Using the MD5 hash of the representations, we determined that this set had 8,859 duplicate representations. With the trimmed URI and the MD5 hash of the entity, we can compare the identifiers and the returned entities for duplication. 


\begin{equation}
\label{acc}
\text{Accuracy} = \frac{\text{True Positives} + \text{True Negatives}}{\text{Number of Classifications}}
\end{equation}

\begin{equation}
\label{fmeas}
\text{F-Measure} = 2 * \frac{\text{Precision} * \text{Recall}}{\text{\text{Precision} + \text{Recall}}}
\end{equation}

For each of the 19,522 URIs in the frontier and their associated entity hash values, we determined the trimmed URI string and the duplications of URIs in the frontier and the number of duplicate URIs that also had a duplicate entity body (Table \ref{dupesTable}). We calculated the accuracy (Equation \ref{acc})\footnote{Accuracy is defined as the number of correctly classified instances divided by the test set size (Equation \ref{acc}). F-Measure extends accuracy to consider the harmonic mean of precision and recall (Equation \ref{fmeas}).} of each trim policy using the number of URIs with the same entity hash and URI as a true positive (TP), the number of URIs that had neither a duplicate URI nor a duplicate entity body as a true negative (TN), and the set of all positives and negatives (P + N) as the total number of URIs (19,522).

\begin{table}
\center
\begin{tabular}{l|r|r||r}
Trim Type    & \begin{tabular}[c]{@{}l@{}}URI \\ Duplicates\end{tabular} & \begin{tabular}[c]{@{}l@{}}URI and \\ Entity\\ Duplicates\end{tabular} & Accuracy \\ \hline\hline
No Trim      & 6,469                        & 4,684                                                                & 0.68                                                                            \\ \hline
Origin Trim  & 7,078                      & 4,749                                                                & 0.68                                                                            \\ \hline
Base Trim    & 10,359                  & 5,191                                                                & 0.56                                                                            \\ \hline
Session Trim & 8,159                         & 4,921                                                                & 0.64                                                                            \\ \hline
HTTP Trim    & 7,315                        & 4,868                                                                    & 0.67                                                                            \\ \hline
\end{tabular}
\caption{Detected duplicate URIs, entity bodies, and the overlap between the two using the five URI string trimming policies.}
\label{dupesTable}
\end{table}

The \emph{Base Trim} and \emph{No Trim} policies had identical accuracy ratings (0.68). The \emph{Base Trim} policy identified the most URI duplicates, and is used to determine the overlap between the Heritrix and PhantomJS frontiers.

Using the \emph{Base Trim} policy to only consider the host and path (e.g., \url{http://pubads.g.doubleclick.net/gampad/ads}) of the PhantomJS and Heritrix frontiers, PhantomJS identifies 376,578 URI-Rs added to the frontier, 199,761 (55\%) of which are duplicates of the discovered URIs. If we consider only the host and path of the PhantomJS URIs, the Euler Diagram of PhantomJS and Heritrix frontiers is more evenly matched (Figure \ref{fuzzy}).

\subsection{Deferred vs. Non-Deferred Crawls}
\label{deferVnon}
To isolate the impact of resources with deferred representations on crawl performance, we manually classified 200 URI-Rs from our set of 10,000 URI-Rs as having deferred representations and another 200 as having non-deferred representations. We crawled each of the deferred and non-deferred sets of URI-Rs with PhantomJS and Heritrix. 

During the crawl of the non-deferred set, PhantomJS crawled $t_{URI}$=0.255 URIs/s while Heritrix crawled $t_{URI}$=1.34 URIs/s, 5.25 times faster than PhantomJS. Heritrix uncovered 1,044 URI-Rs to add to the frontier, while PhantomJS discovered 403 URI-Rs to add to the frontier. This phenomenon of Heritrix having a larger frontier than PhantomJS is due to Heritrix's policy of looking into the JavaScript files to extract URIs found in the code -- the URI-Rs discovered by Heritrix are top-level domains listed in the JavaScript that may be used to construct URIs at run time (e.g., appending a username or timestamp to the URI) or not used by JavaScript at all (e.g., a URI that exists in un-executed code). 

During the crawl of the deferred set, PhantomJS crawled $t_{URI}$=0.5 URIs/s. Heritrix ran $t_{URI}$=12.56 URIs/s, 25.12 times faster than PhantomJS. Heritrix added 3,206 URIs to the frontier, while PhantomJS added 3,436 URIs to the frontier. PhantomJS adds more URIs to the frontier despite Heritrix's introspection on the JavaScript of each crawl target. This result is due to PhantomJS's execution of JavaScript on the client.

We observe that the PhantomJS frontier outperforms the Heritrix frontier during the deferred crawl. Heritrix crawls URIs faster than PhantomJS on each of the deferred and non-deferred crawls, but far exceeds the speed of PhantomJS during the deferred crawl.

\section{Classifying Representations}
\label{classifying}

In practice, archival crawlers such as Heritrix would be able to identify URI-Rs that have low archivability in real-time. Heritrix currently does not have such an automatic capability. Archive-It, for example, uses a manually curated list of URIs that have deferred representatiosn and uses Umbra \cite{umbra} to crawl them.

The ability to determine the archivability of a resource will allow Heritrix to assign the URI-R to either the faster, traditional Heritrix crawler or the slower,  PhantomJS (or other JavaScript-enabled crawler). By enabling this two-tiered approach to crawling, the archival crawlers can achieve maximum performance by utilizing the heavy-duty JavaScript-capable crawlers for only those that need it. However, this approach requires the ability to, in real-time, recognize or predict a deferred representation.

Even though our goal is to detect whether or not representations are dependent on JavaScript, the simple presence of JavaScript is not a sufficient indicator of a deferred representation. In our set of URI-Rs, the resources with deferred representations had, on average, 21.98 embedded script tags or files, while the resources with non-deferred representations had 5.3 script tags or files. Of those resources with deferred representations, 84.1\% had at least one script tag, while 49.5\% of the non-deferred representations had at least one script tag. Because of the ubiquity of JavaScript in both deferred and non-deferred representations, we opted for a more complex feature vector to represent the features of the representations.

In an effort to predict whether or not a representation would be deferred, we constructed a feature vector of DOM attributes and features of the embedded resources. We used Weka \cite{weka} to classify the resources on subsets of the feature vectors to gauge their performance. We extracted the following feature vector:
\begin{enumerate}
\firstitem{ \textbf{Ads}: Using a list of known advertisement domains, we determined whether or not a representation would load an ad based on DOM and JavaScript analysis.}
\item \textbf{Script Tags}: We counted the number of script tags with JavaScript, both in files and embedded code.
\item \textbf{Interactive Elements}: We counted the number of DOM elements that have JavaScript events attached to them (e.g., \texttt{onclick}, \texttt{onload}). 
\item \textbf{Ajax (in JavaScript)}: To estimate the number of Ajax calls (e.g., \texttt{\$.get()}, \texttt{XmlHttpRequest}) we counted the number of occurrences of Ajax requests in the embedded external and independent JavaScript files.
\item \textbf{Ajax (in HTML)}: To estimate the number of Ajax calls (e.g., \texttt{\$.get()}, \texttt{XmlHttpRequest}) we counted the number of occurrences of Ajax requests in Script tags embedded in the DOM.
\item \textbf{DOM Modifications}: We counted the number of times JavaScript made a modification of the DOM (e.g., via the \texttt{appendChild()} function) to account for DOM modifications after the initial page load.
\item \textbf{JavaScript Navigation}: We counted the occurrences of JavaScript redirection and other navigation functions (e.g., \texttt{window.location} calls).
\item \textbf{JavaScript Storage}: We count the number of JavaScript references to storage elements on the client (e.g., cookies) as an indication of client-controlled state.
\item \textbf{Found, Same Domain}: Using PhantomJS, we counted the number of embedded resources originating from the URI-R's top level domain (TLD) that were successfully dereferenced (i.e., returned an HTTP 200).
\item \textbf{Missed, Same Domain}: Using PhantomJS, we counted the number of embedded resources originating from the URI-R's TLD that were not successfully dereferenced (i.e., returned a class HTTP 400 or 500).
\item \textbf{Found, Different Domain}: Using PhantomJS, we counted the number of embedded resources originating outside of the URI-R's TLD that were successfully dereferenced (i.e., returned an HTTP 200).
\item \textbf{Missed, Different Domain}: Using PhantomJS, we counted the number of embedded resources originating outside of the URI-R's TLD that were unsuccessfully dereferenced (i.e., a class 400 or 500 HTTP response).
\end{enumerate}
\vskip -2mm

We manually sampled 440 URI-Rs (from our collection of 10,000, including the same 400 from Section \ref{deferVnon}) and classified the representations as deferred or non-deferred, with 200 training and 20 test URI-Rs for each based on whether or not their representations were dependent upon JavaScript. 


Using PhantomJS, we collected the 12 features required for a feature vector for each of our 440 URI-Rs. Using Weka, we ran each classifier on the feature vectors. Rotation Forests \cite{rotationforest} performed the best of any of the standard Weka classifiers for any of our datasets.


We used three subsets of the feature vector to investigate the best method of predicting deferred representations. We selected attributes 1-8 to represent DOM features. We selected attributes 9-12 as embedded resource attributes (the attributes we extract if we load and monitor the embedded resources). Together, attributes 1-12 make up the entire dataset. We use the feature sets to train and test our classifier via 10-fold cross validation. We use the same three data subsets and provide a confusion matrix of each set including the entire feature vector (Table \ref{cmAll}), resource feature vector (Table \ref{cmResource}), and DOM feature vector (Table \ref{cmDOM}).

\begin{table}
\centering\begin{tabular}{cll}
\textbf{Actual} & \multicolumn{2}{c}{ \textbf{Predicted Classification}}                           \\
\textbf{Classification}   &          Deferred       &           Non-Deferred             \\ \cline{2-3} 
       Deferred           & \multicolumn{1}{|r}{182} & \multicolumn{1}{|r|}{38} \\ \cline{2-3} 
       Non-Deferred       & \multicolumn{1}{|r}{58} & \multicolumn{1}{|r|}{166} \\ \cline{2-3} 
\end{tabular}
 \caption{Confusion matrix for the entire feature vector (F-Measure = 0.791).}
  \label{cmAll}
\end{table}

\begin{table}
\centering\begin{tabular}{cll}
\textbf{Actual} & \multicolumn{2}{c}{ \textbf{Predicted Classification}}                           \\
\textbf{Classification}   &          Deferred       &           Non-Deferred             \\ \cline{2-3} 
       Deferred           & \multicolumn{1}{|r}{179} & \multicolumn{1}{|r|}{41} \\ \cline{2-3} 
       Non-Deferred       & \multicolumn{1}{|r}{47} & \multicolumn{1}{|r|}{173} \\ \cline{2-3} 
\end{tabular}
 \caption{Confusion matrix for the resource features (features 9-12 of the vector; F-Measure = 0.844).}
  \label{cmResource}
\end{table}

\begin{table}
\centering\begin{tabular}{cll}
\textbf{Actual} & \multicolumn{2}{c}{ \textbf{Predicted Classification}}                           \\
\textbf{Classification}   &          Deferred       &           Non-Deferred             \\ \cline{2-3} 
       Deferred           & \multicolumn{1}{|r}{168} & \multicolumn{1}{|r|}{52} \\ \cline{2-3} 
       Non-Deferred       & \multicolumn{1}{|r}{41} & \multicolumn{1}{|r|}{179} \\ \cline{2-3} 
\end{tabular}
 \caption{Confusion matrix for the DOM features (features 1-8 of the vector; F-Measure = 0.806).}
  \label{cmDOM}
\end{table}

The accompanying statistics for the classifications are shown in Table \ref{classStats}. With only the DOM features, the test set is accurately classified representations as deferred or non-deferred 79\% of the time. If we combine the DOM and resource feature sets to create the full feature set, we can correctly classify representations 81\% of the time. 

\begin{center}
\begin{table*}[ht]
\center
\begin{tabular}{l|l|r|r|r|r}
Features                                                                                & Classification & \multicolumn{1}{l}{\begin{tabular}[c]{@{}l@{}}Accuracy\end{tabular}} & \multicolumn{1}{l}{\begin{tabular}[c]{@{}l@{}}F-measure\end{tabular}} & \multicolumn{1}{l}{\begin{tabular}[c]{@{}l@{}}Precision\end{tabular}} & \multicolumn{1}{l}{\begin{tabular}[c]{@{}l@{}}Recall\end{tabular}} \\
\hline
\hline
\multirow{2}{*}{\begin{tabular}[c]{@{}l@{}}DOM \\ Features Only \end{tabular}}         & Deferred       & \multirow{2}{*}{79\%}                                                                 & \multirow{2}{*}{79\%} & 78\%                                                                                   & 81\%                                                                                \\
                                                                                        & Non-deferred   &                                                                                       &  & 76\%                                                                                   & 80\%                                                                                \\
\hline
\multirow{2}{*}{\begin{tabular}[c]{@{}l@{}}DOM \& Resource \\ Features\end{tabular}} & Deferred       & \multirow{2}{*}{81\%}                                                                 & \multirow{2}{*}{82\%} & 79\%                                                                                   & 81\%                                                                                \\
                                                                                        & Non-deferred   &                                                                                       &  & 90\%                                                                                   & 80\% \\               
\hline                                                               
\end{tabular}

 \caption{Classification success statistics for DOM-only and DOM and Resource feature sets.}
  \label{classStats}
\end{table*}
\end{center}

After a URI is dereferenced and a representation is returned, we can determine whether or not the representation is deferred with 79\% accuracy. If we also dereference the URIs for the embedded resources and monitor the HTTP status codes, we can increase, albeit minimally, the accuracy of the prediction to 81\% of the time. However, crawling with PhantomJS is much more expensive when executed properly. Due to this minimal improvement and much higher cost to measure, the feature extraction will be limited to the DOM classification. With a negligible impact on performance, our classifier is able to identify deferred representations using the DOM crawled by Heritrix with 79\% accuracy.


\section{Two-Tiered crawling}
\label{frame}

To benefit from the increased crawl frontier size of PhantomJS while maintaining the performance of Heritrix, we propose a tiered crawling approach in which PhantomJS is used to crawl only resources with deferred representations. 
A tiered approach to crawling would allow an archive to simultaneously benefit from the frontier size of PhantomJS and the speed of Heritrix. 
Table \ref{finalStatTable} provides a summary of the extrapolated crawl speed and discovered frontier size of each crawler. While the test environment used a single system, a production environment should expect to see performance improvements with additional resources. PhantomJS crawls are not run in parallel, and additional nodes for PhantomJS threads will further improve performance.

\begin{table}[h]
\begin{tabular}{l|r|r|r}
Crawl Strategy                                                                     & \begin{tabular}[c]{@{}l@{}}Crawl Time\\ (hrs)\end{tabular} & \begin{tabular}[c]{@{}l@{}}Crawl Rate\\ ($t_{URI}$)\end{tabular} & \begin{tabular}[c]{@{}l@{}}Frontier Size\\ ($|F|$)\end{tabular} \\ 
\hline
\hline
wget                                                                               & 416.16                                                     & 0.864                                                           & 129,443                                                        \\  
\hline
Heritrix                                                                           & 407.53                                                     & 2.065                                                           & 302,961                                                        \\ 
\hline
PhantomJS                                                                          & 8,684.38                                                   & 0.170                                                           & 531,484                                                        \\ 
\hline
\begin{tabular}[c]{@{}l@{}}Heritrix +\\  PhantomJS\end{tabular}                    & 9,100.54                                                   & 0.152                                                           & 537,609                                                        \\ 
\hline
\begin{tabular}[c]{@{}l@{}}Heritrix + \\ PhantomJS \\ with Classifier\end{tabular} & 6,495.23                                                   &  0.196  & 458,815                                                        \\ 
\hline
\end{tabular}
\caption{A summary of \emph{extrapolated} performance (based on our calculations) of single- and two-tiered crawling approaches.}
\label{finalStatTable}
\end{table}

We have described the operation of crawls with wget, Heritrix, and PhantomJS in Sections \ref{clock} and \ref{deferVnon} with wget serving as a baseline to which Heritrix and PhantomJS can be compared but wget is not part of the archival workflow we investigate. To reiterate, Heritrix crawls much more quickly than PhantomJS, while PhantomJS discovers many more embedded resources required to properly construct a representation. Optimally during a crawl, Heritrix would dereference a URI-R and run the resulting DOM through the classifier to determine whether or not the representation will be deferred (with 79\% accuracy, as discussed in Section \ref{classifying}). If the representation is predicted to be deferred, PhantomJS should also be used to crawl the URI-R and add the newly discovered URI-Rs to the Heritrix frontier.

Heritrix should be used to crawl all URI-Rs in the frontier because the DOM is required to classify a representation as deferred. Since Heritrix is the fastest crawler, it should be used to dereference the URI-Rs in the frontier and retrieve the DOM of the resource for classification. Subsequently, only if the representation is classified as deferred will PhantomJS be used to crawl the resource to ensure the maximum amount of embedded resources are retrieved.

In a naive two-tiered crawl strategy that will discover the most embedded URI-Rs and create the largest frontier, Heritrix and PhantomJS should both crawl each URI-R regardless of whether the representation can be classified as deferred or non-deferred. This creates a crawl that is expected to be 13.5 times slower than simply using Heritrix, but is expected to discover 1.77 times more URI-Rs than using only Heritrix. This would ensure that 100\% of all resources with deferred representations would be crawled with both Heritrix and PhantomJS. However, we want to limit the use of PhantomJS to minimize the performance impacts it has on the crawl speed.

If we include the classifier to predict when PhantomJS should be used or when Heritrix will be a suitable tool, the two-tiered approach is expected to run 10.5 times slower and is expected to discover 1.5 times more URI-Rs than only Heritrix. This crawl policy balances the trade-offs between speed and larger frontier size by using the classifier to indicate when to use PhantomJS to crawl resources with deferred representations.

To validate this expected calculation, we classified our 10,000 URI-R dataset, which produced 5,187 URI-Rs classified as having deferred representations, and 4,813 as having non-deferred representations. We used PhantomJS to crawl the URI-Rs classified as deferred, and only Heritrix to crawl the URI-Rs classified as non-deferred. The results of the crawls are detailed in Table \ref{classifyCrawl}.

\begin{table}[h]
\begin{tabular}{llrrr}
Crawler                     & URI-R Set                         & \begin{tabular}[c]{@{}l@{}}Seed \\ Size\end{tabular} & \begin{tabular}[c]{@{}l@{}}Frontier \\ Size\end{tabular} & \begin{tabular}[c]{@{}l@{}}Crawl \\ Time (hrs)\end{tabular}                                                                             \\ \hline
\multicolumn{1}{|l|}{P}   & \multicolumn{1}{l|}{Deferred}     & \multicolumn{1}{r|}{5,187}                           & \multicolumn{1}{r|}{311,903}                             & \multicolumn{1}{r|}{84.9}  \\ \hline
\multicolumn{1}{|l|}{H} & \multicolumn{1}{l|}{Non-deferred} & \multicolumn{1}{r|}{4,813}                           & \multicolumn{1}{r|}{124,728}                             & \multicolumn{1}{r|}{23.6}   \\ \hline
\multicolumn{1}{|l|}{H} & \multicolumn{1}{l|}{Deferred}     & \multicolumn{1}{r|}{5,187}                           & \multicolumn{1}{r|}{171,499}                             & \multicolumn{1}{r|}{26.7}   \\ \hline
\multicolumn{1}{|l|}{P}   & \multicolumn{1}{l|}{All URI-Rs}   & \multicolumn{1}{r|}{10,000}                          & \multicolumn{1}{r|}{438,388}                             & \multicolumn{1}{r|}{686}\\ \hline
\multicolumn{1}{|l|}{H} & \multicolumn{1}{l|}{All URI-Rs}   & \multicolumn{1}{r|}{10,000}                          & \multicolumn{1}{r|}{275,234}                             & \multicolumn{1}{r|}{48.3}\\ \hline
\multicolumn{1}{|l|}{Two-tier} & \multicolumn{1}{l|}{All URI-Rs}   & \multicolumn{1}{r|}{10,000}                          & \multicolumn{1}{r|}{399,202}                             & \multicolumn{1}{r|}{133} \\ \hline
\end{tabular}
\caption{A simulated two-tiered crawl showing that the frontier sizes can be optimized while mitigating the performance impact of PhantomJS's (P) crawl speed vs Heritrix's (H).}
\label{classifyCrawl}
\end{table}

In this table, we show that PhantomJS creates a frontier of 438,388, 1.6 times larger than that of Heritrix. However, PhantomJS crawls 14 times slower than Heritrix. If we perform a tiered crawl in which PhantomJS is responsible for crawling only deferred representations, we can crawl 5.2 times faster than using only PhantomJS (but 2.7 times slower than the Heritrix-only approach) while creating a frontier 1.8 times larger than using only Heritrix. As a result, we can maximize the frontier size, mitigate the impacts of JavaScript on crawling, and mitigate the impact of the reduced crawl speeds when using a tiered crawling approach.

\section{Conclusions}
\label{conclusion}

In this paper, we measured the differences in crawl speed and frontier size of wget, PhantomJS, and Heritrix. While PhantomJS was the slowest crawler, it provided the largest crawl frontier due to its ability execute client-side JavaScript to discover URIs missed by Heritrix and wget. Heritrix was the fastest crawler. We also proposed a tiered approach to crawling in which a classifier determines whether to crawl a resource with PhantomJS to reap the URI discovery benefits of the specialized crawler where appropriate.

This work lays the foundation for a two-tiered crawling approach and helps predict the performance of future archival workflows. We know that PhantomJS finds 19.70 more embedded resources per URI and Heritrix runs 12.13 times faster than PhantomJS, meaning the crawler should avoid crawling URI-Rs with non-deferred representations to maintain an optimal performance trade-off. We understand that PhantomJS is required to discover the embedded resources needed to complete a deferred representation that Heritrix cannot discover. This has a performance detriment to run time, but offers a benefit of more complete mementos and a larger frontier for crawling. We also found that 53\% of URIs discovered by PhantomJS are duplicates if we remove session-specific URI parameters. 

Using DOM features we can accurately predict deferred and non-deferred representations 79\% of the time. Using this classification, deferred representations can be crawled by PhantomJS to ensure all embedded resources are added to the crawl frontier.

If using a multi-tiered approach to crawling, archives can leverage the benefits of PhantomJS and Heritrix simultaneously. That is, using a deferred representation classifier, archives can use PhantomJS for deferred representations and Heritrix for non-deferred representations. Using a tiered crawling approach, we showed that crawls will run 5.2 times faster than using only PhantomJS, create a frontier 1.8 times larger than using only Heritrix. This crawl strategy mitigates the impact of JavaScript on archiving while also mitigating the reduced crawl speed of PhantomJS.

Our future work will include a framework for archiving deferred representations, along with a measurement of the archival improvement when implementing a deferred representation crawler.

\section{Acknowledgments}
This work supported in part by the NEH (HK-50181).

\bibliographystyle{abbrv}

\end{document}